\documentclass{article}
\usepackage{spconf,amsmath,graphicx}
\usepackage{epsfig} 
\usepackage{floatflt} 
\usepackage{paralist} 
\usepackage{algorithm} 
\usepackage{amsthm}
\usepackage{amssymb}
\usepackage{listings}    
\usepackage{xcolor}
\usepackage{hyperref}
\usepackage{indentfirst}
\usepackage{booktabs}
\usepackage{subfigure}
\usepackage{caption}
\usepackage{calligra}
\usepackage{bm}
\usepackage{fancyhdr}
\usepackage{float}
\usepackage{pifont}
\usepackage{colortbl} 
\usepackage{MnSymbol}%
\usepackage{amsbsy}

\usepackage{enumitem}

\def\a{\mathbf{a}}

\def\c{\mathbf{c}}
\def\x{\mathbf{x}}

\def\y{\mathbf{y}}

\def\p{\mathbf{p}}

\def\s{\mathbf{s}}

\def\N{\mathcal{N}}
\def\V{\mathcal{V}}

\newcommand{\mypar}[1]{{\bf #1.}}
\theoremstyle{definition}

\DeclareMathOperator{\Adj}{A}

\DeclareMathOperator{\Pj}{P}

\DeclareMathOperator{\X}{X}

\newcommand{\R}{\ensuremath{\mathbb{R}}}

\title{PCT: Large-Scale 3D Point Cloud Representations via Graph Inception Networks with Applications to Autonomous Driving }

\name{Siheng Chen, Sufeng Niu, Tian Lan, Baoan Liu}
\address{}

\begin{document}
%
\maketitle
\begin{abstract}
We present a novel graph-neural-network-based system to effectively represent large-scale 3D point clouds with the applications to autonomous driving. Many previous works studied the representations of 3D point clouds based on two approaches, voxelization, which causes discretization errors and learning, which is hard to capture huge variations in large-scale scenarios.  In this work, we combine voxelization and learning:  we discretize the 3D space into voxels and propose novel graph inception networks to represent 3D points in each voxel. This combination makes the system avoid discretization errors and work for large-scale scenarios. The entire system for large-scale 3D point clouds acts like the blocked discrete cosine transform for 2D images; we thus call it the point cloud neural transform (PCT). We further apply the proposed PCT to represent real-time LiDAR sweeps produced by self-driving cars and the PCT with graph inception networks significantly outperforms its competitors.
\end{abstract}

\section{Introduction}
\label{sec:intro}
With the growth of 3D sensing technologies, one can now use a large number of 3D points to precisely represent objects' surfaces and surrounding environments. We call those 3D points a~\emph{3D point cloud}; it has a growing impact on various applications, including autonomous driving, virtual reality and scanning of historical artifacts~\cite{PCL}.  In this paper, we consider the setting of autonomous driving. A self-driving car could use multiple sensors to observe the world, such as LiDARs, cameras and RADARs~\cite{GeigerLU:12}. Among those, LiDARs produce two types of 3D point clouds, real-time LiDAR sweeps and high-precision maps. Both include accurate range information, which are critical to perception and localization systems. We consider both types of point clouds~\emph{large-scale point clouds} because they contain a large number of 3D points and record outdoor environments. 

\begin{figure}[thb]
  \begin{center}
    \begin{tabular}{cc}
    \includegraphics[width=0.36\columnwidth]{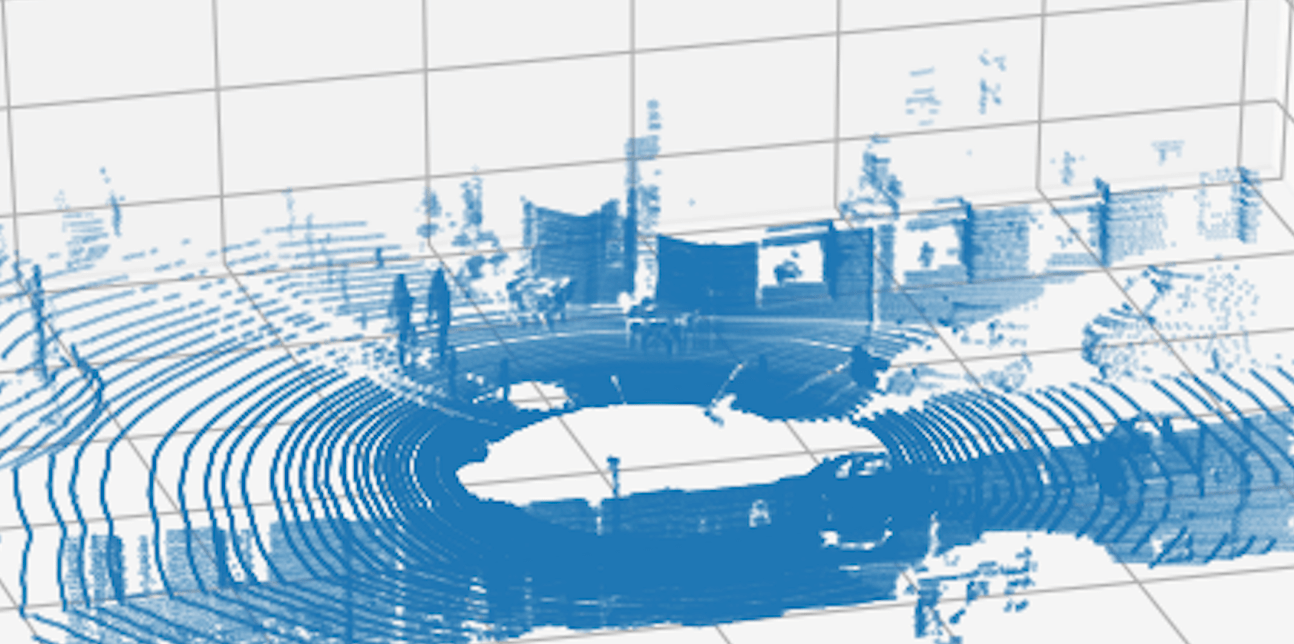}     & \includegraphics[width=0.36\columnwidth]{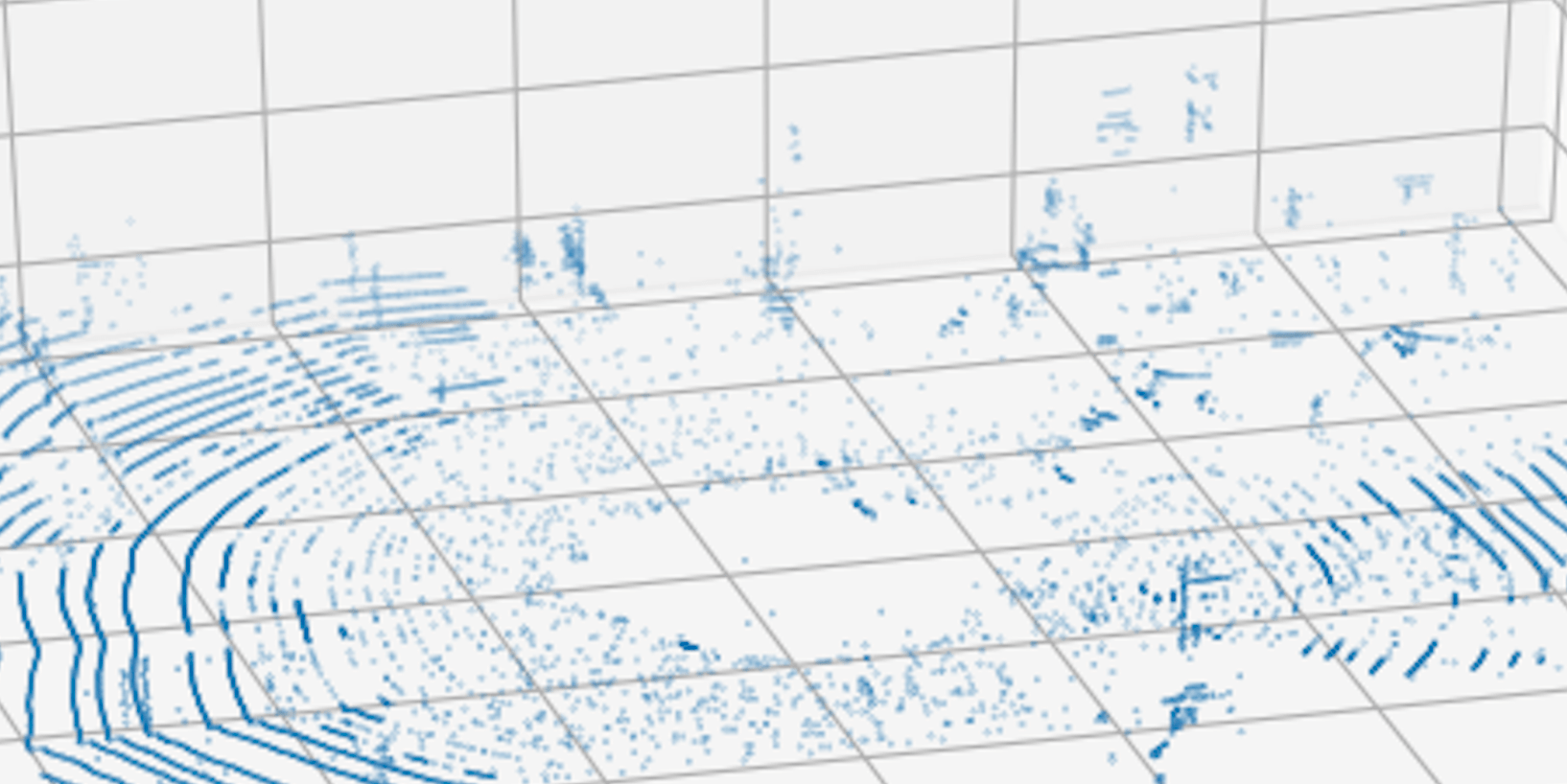}   
    \\
    {\small (a) Original LiDAR sweep.} &  {\small (b) Globally  uniform resampling.}
    \\
        \includegraphics[width=0.36\columnwidth]{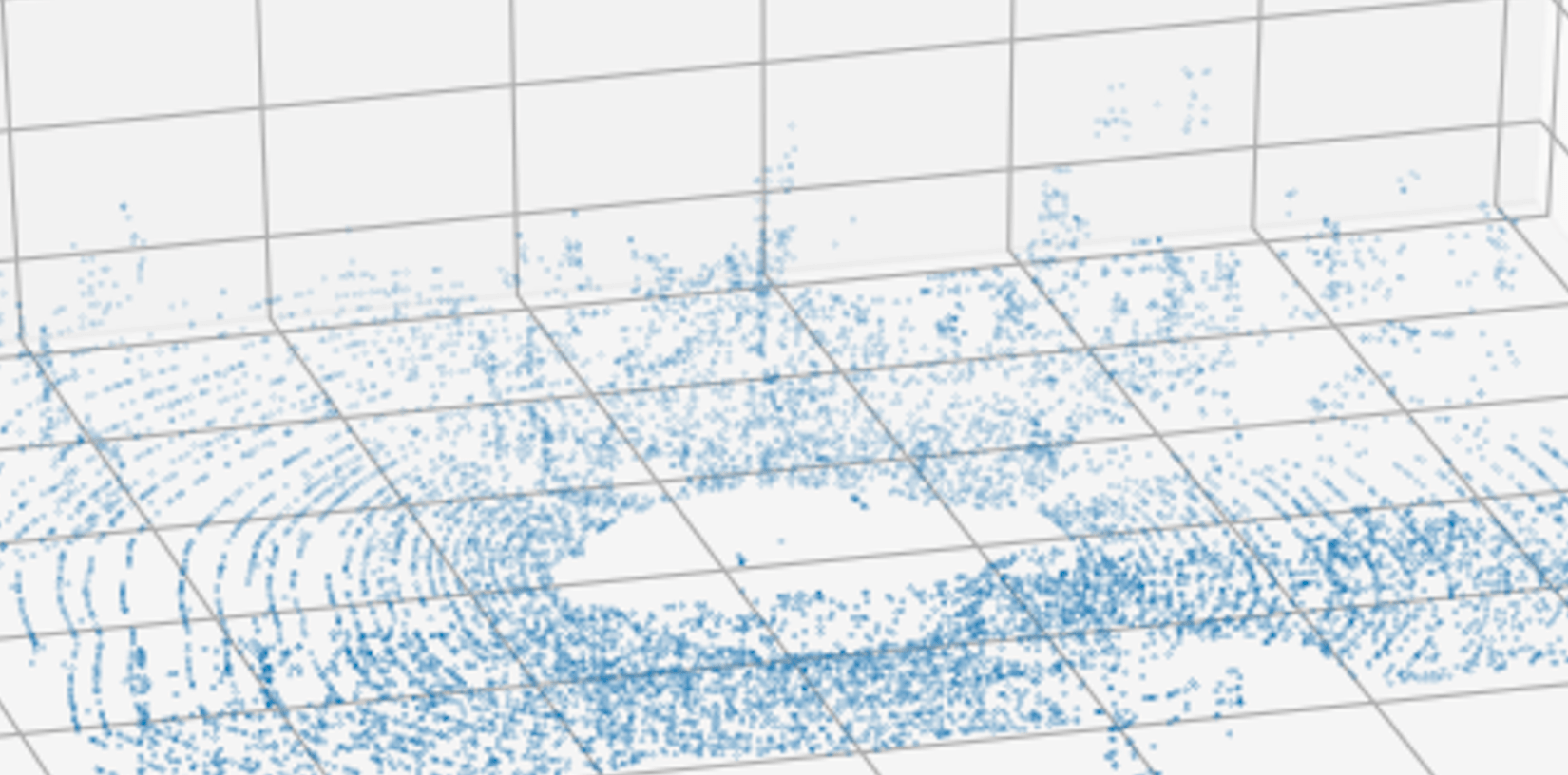}     & \includegraphics[width=0.36\columnwidth]{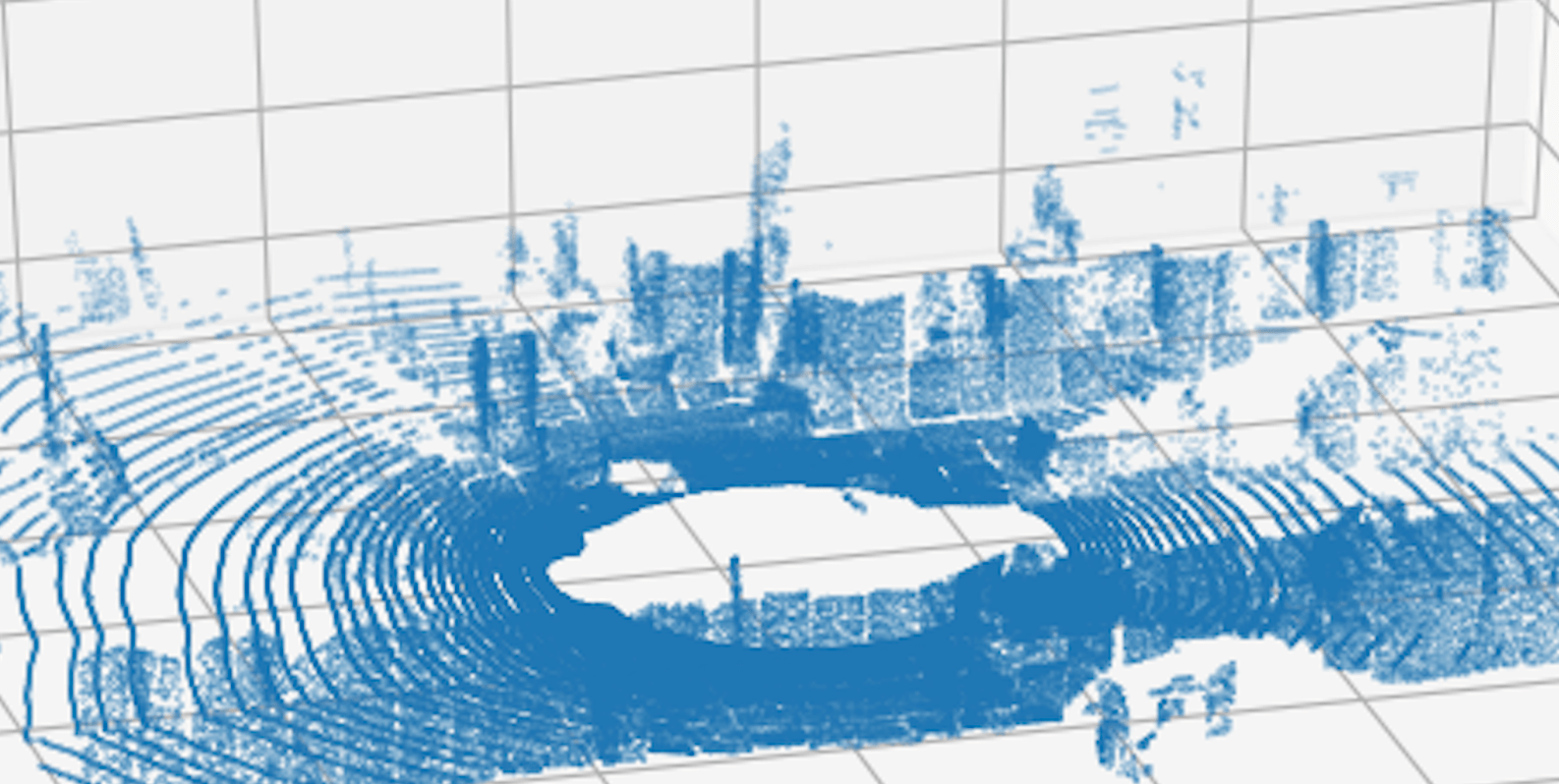}   
    \\
    {\small (c) Octree.} &  {\small (d) PCT: GIN.}
  \end{tabular}
\end{center}
\caption{\label{fig:crown}  Comparison between an original sweep and its reconstructions. All reconstructions use $2.78\%$ of original data. }
\end{figure}

To fully exploit real-time LiDAR sweeps and high-precision maps, we need advanced techniques to handle a series of challenges, including 3D point cloud compression, 3D localization and 3D object detection. A common task shared in those challenges is~\emph{3D point cloud representations}; that is, representing a 3D point cloud in a compact format, such that it is easy to conduct subsequent processing procedures. For 1D time-series, the basic representation is the Fourier transform; for 2D images, it is the discrete cosine transform and 2D wavelet transform~\cite{VetterliKG:12}; for 3D point clouds, Octree partitions the 3D space adaptively and has been an effective representation tool~\cite{SchnabelK:06}; however, Octree represents a 3D point cloud only in the 3D spatial domain and does not fully exploit shapes formed by 3D points.

To represent large-scale 3D point clouds, we propose a novel graph-neural-network-based system, called the~\emph{point cloud neural transform} (PCT). The PCT includes two phases: voxelization, which adopts the standard Octree-like partition  and splits a large-scale space into a series of small-scale spaces (voxels), and voxel-level encoding, which adopts graph neural networks to capture the complicated underlying distributions of 3D points within each voxel. In the phase of voxel-level encoding, we propose novel graph inception networks, which transform 3D points in a voxel to a low-dimensional feature vector and extend to the translation-invariant graph convolution~\cite{NiuCGTSK:18} to the 3D space. Because of the two-phase design, PCT can be considered as the 3D counterpart of the windowed Fourier transform for 1D time series and the blocked discrete cosine transform for 2D images. We further apply the PCT to represent real-time LiDAR sweeps collected by self-driving cars and compare it with the standard Octree-based representations. As a general tool, the PCT can be potentially used to 3D compression, 3D object detection and many others in autonomous driving.  The main contributions of the paper include: (i) we propose the PCT to represent large-scale 3D point clouds; (ii) we propose novel graph inception networks to implement the voxel-level encoding;  (iii) the proposed PCT is applied to represent real-time LiDAR sweeps produced by self-driving cars and outperforms its competitors.

\mypar{Related works}
3D point cloud processing has become important in various 3D imaging and autonomous systems. The topic broadly includes
  compression~\cite{SchwarzPBBCCCKL:19,ThanouCF:16,AnisCO:16}, denoising~\cite{DuanCK:18, ZengCNPY:18},  surface  reconstruction~\cite{GregorskiHJ:00}, feature  extraction~\cite{FengTK:14}, localization~\cite{SpangenbergGR:16}, 3D object detection~\cite{LiangYWU:18} and many others. Here we consider 3D point cloud representations, which mainly based on three approaches, including resampling, voxelization and learning. Resampling represents a 3D point cloud by selecting a subset of 3D points~\cite{ChenTFVK:18}. In many applications, resampling can enhance key geometry information and make subsequent processing both cheaper and more accurate~\cite{GelfandIRL:2003}. Voxelization represents a 3D point cloud by partitioning the 3D space into a series of voxels and use the corresponding voxel center as the proxy for each 3D point, such as regular voxels~\cite{Loop:13}  and Octree~\cite{HornungWBSB:13}; however, they all suffer from discretization errors.  Learning-based models can also be trained to capture the underlying distribution of 3D points. For example,~\cite{EckartKTKK:16}
proposed a probabilistic generative model to model the distribution of 3D point clouds; however, such model is inefficient in inferring parameters; ~\cite{AchlioptasDMG:17} proposed a deep  autoencoder  that directly handles 3D point clouds; ~\cite{YangFST:18, GroueixFKRM:18} introduced a 2D lattice to help decoding.
\section{Methodology}
\label{sec:method}
Similarly to many standard representation problems, the overall goal is to use a low-dimensional feature vector to represent a large-scale 3D point cloud; however, a large-scale 3D point cloud has its own challenges:
(i) variations. 3D points captured in a outdoor environment have huge variations, while the available training data are limited. The representations need to learn rich variations from limited amount of 3D point clouds; (ii) irregularity. 3D points are irregularly and sparsely scattered in the 3D space. The representations need to go beyond  the regular lattices and capture irregular and nonuniform distributions of 3D points; (iii) invariances and equivalences. The representations need to promote basic geometric properties, including permutation-invariance, translation-invariance, scale-equivalent and rotation-equivalence. We propose the point cloud neural transform (PCT) to handle these challenges.

\begin{figure}
  \begin{center}
\includegraphics[width=0.75\columnwidth]{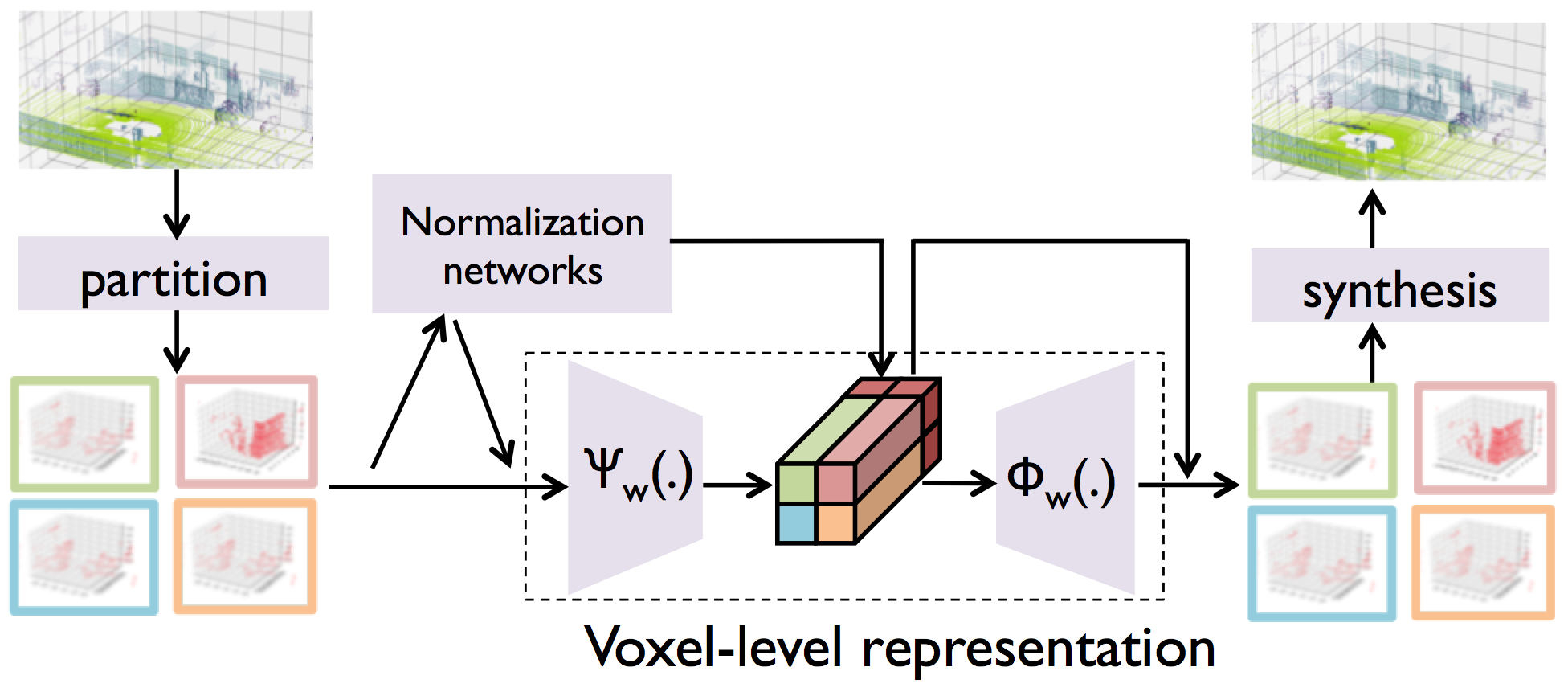}  
  \end{center}
  \caption{\label{fig:two_stage_pipeline}  Proposed point cloud neural transform (PCT). }
\end{figure}

\mypar{Point cloud neural transform}
To handle large variations, we propose a two-stage framework. In the first stage, we partition the 3D space into a series of voxels; in the second stage,  we encode 3D points in each voxel to a few codes; see the entire system in Figure~\ref{fig:two_stage_pipeline}. The intuitions are (i) since the representations in each individual voxel is much more constraint, we are able to specifically and effectively  learn local shapes and patterns; and (ii) since each 3D point cloud produces multiple voxels, we are able to increase the number of training data and potentially train strong voxel-level representations. This two-stage representation is similar to the windowed Fourier transform for 1D time series and the blocked discrete cosine transform for 2D images. To handle irregularity, we propose graph inception networks, which use a graph to capture the underlying distribution of 3D points. We treat 3D points as nodes and connects each point to its neighboring points to formulate a spatial graph. The edge weights reflect irregular and nonuniform distribution of 3D points.  The graph inception networks also ensure permutation and translation-invariances. To handle equivalences, we propose normalization networks to ensure scale and rotation-equivalences.

Mathematically, let $\Pj \in \R^{N \times 3}$ be the matrix representation of this 3D point cloud, whose $i$th row $\p_i = [x_i, y_i, z_i] \in \R^3$ represents the 3D coordinate of the $i$th point. The overall procedures of the PCT are
\begin{subequations}
\label{eq:PCT}
\begin{eqnarray}
\label{eq:partition}
 \{ \Pj^{(i)} \}_{i=1}^M & = &   \rm{partition} \left( \Pj \right),
\\
\label{eq:voxel_encoder}
	\c^{(i)}   & = &  \Psi_w \left( \Pj^{(i)} \right) \in \R^\ell,
\\
\label{eq:voxel_decoder}
	\widehat{\Pj}^{(i)}  & = &  \Phi_w \left( \c^{(i)} \right)
\\
\label{eq:synthesis}
   \widehat{\Pj}   & = & {\rm synthesis} \left(  \{ \widehat{\Pj}^{(i)} \}_{i=1}^M \right).
\end{eqnarray}
\end{subequations}
In~\eqref{eq:partition}, we partition a 3D space into a series of nonoverlapping voxels based on the spatial structure; correspondingly, a large-scale 3D point cloud is partitioned into a series of small-scale 3D point clouds, where $\Pj^{(i)}$ represents the 3D point cloud in the $i$th voxel. In~\eqref{eq:voxel_encoder}, we encode the 3D points in each voxel to a low-dimensional feature vector. In~\eqref{eq:voxel_decoder}, we decode a low-dimensional feature vector back to the 3D coordinates. In~\eqref{eq:synthesis}, we concatenate the 3D coordinates in all the voxels and reconstruct a 3D point cloud.  

\mypar{Partition}
We simply partition the 3D space into equally-spaced nonoverlapping voxels from each of three dimensions. Let each voxel is of size $H, W, D$ along the $X, Y, Z$ axes respectively.  The $(h, w, d)$th voxel represents a 3D space,
\begin{eqnarray}
\label{eq:voxelization}
\V_{h,w, d}  \ = \ \{ (x, y, z)  |  & (h-1) H   \leq x < h H, &
\\ \nonumber
& (w-1) W \leq y < w W, &
\\ \nonumber
& (d - 1 )D  \leq z < d D \}.
\end{eqnarray}
The points inside $\V_{h,w, d}$ form a point cloud  $\Pj_{h,w, d}$. The partition~\eqref{eq:voxelization} contributes to the implement of~\eqref{eq:partition} and~\eqref{eq:synthesis}. The PCT is also compatible to multiscale voxels, such as Octree.

\mypar{Graph inception networks as voxel-level encoder}  We use the graph inception networks (GIN) to implement the voxel-level encoding~\eqref{eq:voxel_encoder}. The operations and weights of GIN are shared across all the voxels. To handle the irregularity, we introduce a spatial graph to capture the distribution of 3D points. For the simplicity, here we consider a $K$-nearest-neighbor graph, where each 3D points connects to its $K$ closest 3D points. We denote the neighboring set for of $i$-th point as $\N^{(K)}_i$. We then use an adjacency matrix $\Adj^{(K)} \in \R^{n \times n}$ to reflect the pairwise connectivity, where $n$ is the number of 3D points in the corresponding voxel. The edge weight between two points $\p_i$ and $\p_j$ is
\begin{equation*}
  \label{eq:adj}
  \Adj^{(K)}_{i,j} = 
  \left\{ 
    \begin{array}{rl}
      e^{- \left\| \p_i - \p_j \right\|_2^2 }, & j \in \N^{(K)}_i;\\
      0, & \mbox{otherwise}.
    \end{array} \right. 
\end{equation*}

\begin{figure}
  \begin{center}
\includegraphics[width=0.55\columnwidth]{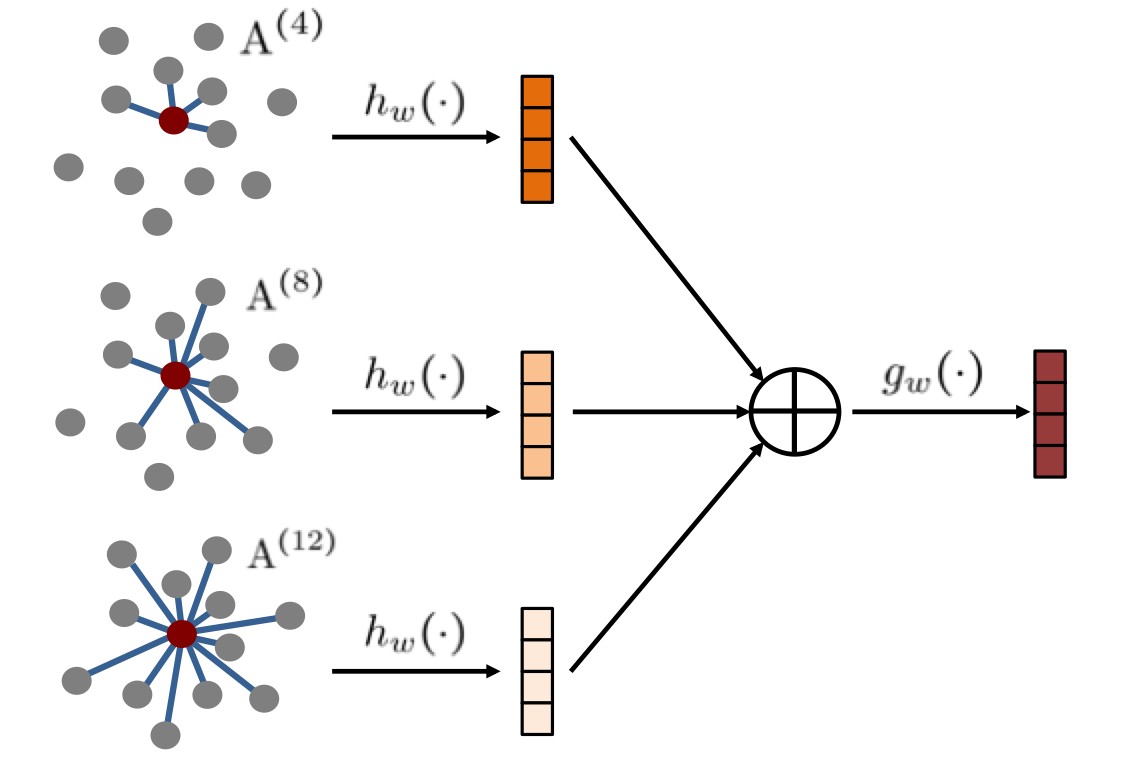}  
  \end{center}
  \caption{\label{fig:gin}  Proposed graph inception convolution: ${\rm conv_{gin}}(\cdot)$. }
\end{figure}

A key issue raised by a graph-based approach is how to choose the number of neighbors $K$. Especially, the point density could vary a lot in various voxels. A fixed $K$ could either capture  limited information or irrelevant information. To solve this issue, we adopt an inceptive-like structure, where we construct a series of $K$-nearest-neighbor graphs with multiple $K$ values. This $K$ value is equivalent to the kernel size of the classical 2D convolution: a larger $K$ indicates a larger reception field.  We thus consider a graph inception convolution to extract features from 3D points. Let $\x_i \in \R^\ell$ be the features of the $i$-th point (the initial feature is $\p_i \in \R^3$). The response is
\begin{eqnarray}
\label{eq:graph_inception_convolution_point}
 \x'_i & = &  g_{w} \left( \Big[ \y^{(k)}_i \Big]_{k \in \mathcal{K}}  \right)
 \\
 & = &  g_{w} \left( \Big[ \sum_{j \in \N^{(k)}_i} h_{w} \left( \Adj^{(k)}_{i,j},  \x_j - \x_i \right)   \Big]_{k \in \mathcal{K}}  \right) \in \R^{\ell'},
\end{eqnarray}
where $h_{w} (\cdot)$ is a standard multilayer perceptron with parameters $w$, $\mathcal{K} $ is a set of $K$ values, $[\cdot]_{k \in \mathcal{K}}$ denotes the concatenation and $g_{w}(\cdot)$ is the inception network that combine responses from multiple graph convolutions. The edge weight $\Adj^{(K)}_{i,j}$ and the difference $\x_j - \x_i$  reflects the relative difference of two points in the original 3D space and the feature space, respectively.  The corresponding matrix representation is
\begin{eqnarray}
\label{eq:graph_inception_convolution}
 \X' = {\rm conv_{gin}}  (\big[ \Adj^{(k)} \big]_{k \in \mathcal{K}} , \X) \in \R^{n \times \ell'}, 
\end{eqnarray}
where $\X \in \R^{n \times \ell}$ is the input feature matrix with the $i$th row vector $\x_i$ in~\eqref{eq:graph_inception_convolution_point} and $\X'$ is the output feature matrix with the $i$th row vector $\x'_i$.  Since we only consider relative differences, the graph convolution is translation-invariant; that is,
$$
 {\rm conv_{gin}}  (\X + {\bf 1}_n \a^T) \ = \  {\rm conv_{gin}}  (\X),
$$
holds for arbitrary $\a \in \R^\ell$, where ${\bf 1}_n \in \R^n$ is a all-one vector. We call~\eqref{eq:graph_inception_convolution}~\emph{graph inception convolution}; see Figure~\ref{fig:gin}. 

After a series of graph inception convolution, we obtain deep point-wise features. To produce voxel-level features, we average the features across all the points.  Let $\X \in \R^{n \times \ell}$ be the final point-wise feature matrix, $\c \ = \   {\rm agg_{mean}} (\X) =   {\X^T {\bf 1}_{n}}/{n}    \in \R^\ell$. Since we aggregate along the point dimension, the final voxel-level features are permutation invariant. Comparing to the max-aggregation~\cite{AchlioptasDMG:17}, the mean-aggregation allows all points contributes to the code, which preserves richer information for reconstruction. 

The voxel-level encoder $\Psi_{w}(\cdot)$ is thus a combination of graph inception convolutions and a mean aggregation. For example, the encoder of the $i$th voxel with a single layer graph inception convolution is
\begin{equation*}
\c^{(i)}  \ = \  \Psi_{w}(\Pj^{(i)}) =  {\rm agg_{mean}} \left( \rho( {\rm conv_{gin}}  (\big[ \Adj^{(k)} \big]_{k \in \mathcal{K}} , \Pj^{(i)}) ) \right),
\end{equation*}
where $\rho(\cdot)$ is the nonlinear activation, such as ReLU. In our experiments, we use $3$-layer graph inception convolutions.

\mypar{Graph inception convolution is a 3D convolution}
Here we consider graph inception convolution from a different perspective. We can represent each 3D point as a delta function in the 3D space. A 3D point cloud is then a train of delta functions; that is,
$$
\s(\p) = \sum_{j}\delta (\p - \p_j),
$$
where $\p, \p_j \in \R^3$. Let a 3D convolution be $h(\cdot): \R^3 \rightarrow \R$. The response is then
\begin{eqnarray*}
y(\p) & = & \int_\tau h(\p - \tau) \s(\tau) d \tau
\\
& = &  \int_\tau h(\p - \tau) \sum_{j}\delta (\tau - \p_j) d \tau
\\
& = &  \sum_{j} h(\p - \p_j).
\end{eqnarray*}
This is equivalent to the graph inception convolution in~\eqref{eq:graph_inception_convolution_point}. In other words, we aim to learn a kernel function $h(\cdot)$ through neural networks. The number of neighbors in the graph reflects the size of the reception field of a kernel function. This kernel function is learnt in the 3D continuous space and operates in the graph domain. The nature of a 3D convolution indicates the properties of translation invariance and weight sharing in the 3D space; it also ensures that the same kernel function works for various graph topologies. Once the neighboring points form a same shape pattern, we will construct the same local graph topology and obtain the same response.

\mypar{Fully connected layers as voxel-level decoder}
To design a voxel-level decoder $\Phi_{w}(\cdot)$ in~\eqref{eq:voxel_decoder}, we cannot use graph-based approaches, because the only information that decoder can access is the code given by the encoder and the graph information no longer exists. We consider two approaches to design a voxel-level decoder. The first approach is based on fully-connected layers, which use more trainable parameters and work better in practice~\cite{AchlioptasDMG:17}; the second approach is based on FoldingNet~\cite{YangFST:18, GroueixFKRM:18}, which considers that points are warped from a 2D map. 

\begin{table*}
  \footnotesize
  \begin{center}
   \begin{tabular}{@{}
>{\columncolor[RGB]{242,229,229}}c
>{\columncolor[RGB]{242,229,229}}c
>{\columncolor[RGB]{242,229,229}}c
>{\columncolor[RGB]{229,229,242}}c
>{\columncolor[RGB]{229,229,242}}c
>{\columncolor[RGB]{229,242,229}}c
>{\columncolor[RGB]{229,242,229}}c
>{\columncolor[RGB]{229,242,229}}c
>{\columncolor[RGB]{229,242,229}}c
>{\columncolor[RGB]{229,242,229}}c@{}}
      \toprule
      \multicolumn{3}{>{\columncolor[RGB]{242,229,229}}c}{\bf  Encoder: GIN}  & 
      \multicolumn{2}{>{\columncolor[RGB]{229,229,242}}c}{\bf Decoder}  
&
	   \multicolumn{5}{>{\columncolor[RGB]{229,242,229}}c}{\bf Metrics}  
 \\
         \midrule \addlinespace[1mm] 
\multicolumn{1}{>{\columncolor[RGB]{242,229,229}}c}{\bf Inception}    & 
       \multicolumn{1}{>{\columncolor[RGB]{242,229,229}}c}{\bf Mean-Agg}   
&
       \multicolumn{1}{>{\columncolor[RGB]{242,229,229}}c}{\bf Norm}  
& 
       \multicolumn{1}{>{\columncolor[RGB]{229,229,242}}c}{\bf FC}
&  
      \multicolumn{1}{>{\columncolor[RGB]{229,229,242}}c}{\bf FoldingNet}   
&       {\bf EMD}   
&       {\bf CD}  
&       {\bf Mean}    
&       {\bf Variance} 
&       {\bf MSE}    
       \\
      \midrule \addlinespace[1mm] 
  &    & & $\pmb{\checkmark}$ &  & $2.947 \times 10^{1}$ & $2.689 \times 10^{-2}$ &  $4.531 \times 10^{-2}$ & $1.832 \times 10^{-3}$ & $3.933 \times 10^{-3}$
\\
$\pmb{\checkmark}$  &    & & $\pmb{\checkmark}$ &  & $2.552 \times 10^{1}$ & $2.342 \times 10^{-2}$ &  $4.220 \times 10^{-2}$ & $1.842 \times 10^{-3}$ & $3.668 \times 10^{-3}$
\\
$\pmb{\checkmark}$  &    & &  & $\pmb{\checkmark}$  & $2.707 \times 10^{1}$ & $2.489 \times 10^{-2}$ &  $4.354 \times 10^{-2}$ & $1.930 \times 10^{-3}$ & $3.871 \times 10^{-3}$
\\
$\pmb{\checkmark}$ &  $\pmb{\checkmark}$   & & $\pmb{\checkmark}$ &  & $2.429 \times 10^{1}$ & $2.296 \times 10^{-2}$ &  $4.164 \times 10^{-2}$ & $1.716 \times 10^{-3}$ & $3.489 \times 10^{-3}$
\\
$\pmb{\checkmark}$ &  $\pmb{\checkmark}$   & &  & $\pmb{\checkmark}$  & $2.606 \times 10^{1}$ & $2.464 \times 10^{-2}$ &  $4.321 \times 10^{-2}$ & $1.792 \times 10^{-3}$ & $3.699 \times 10^{-3}$
\\
$\pmb{\checkmark}$ &  $\pmb{\checkmark}$   & $\pmb{\checkmark}$ & $\pmb{\checkmark}$ &  &${\bf 2.237 \times 10^{1}}$ & ${\bf 1.88 \times 10^{-2}}$ &  ${\bf 3.982 \times 10^{-2}}$ & ${\bf 1.706 \times 10^{-3}}$ & ${\bf 3.335 \times 10^{-3}}$
\\
$\pmb{\checkmark}$ &  $\pmb{\checkmark}$   & $\pmb{\checkmark}$ &  & $\pmb{\checkmark}$  & $2.328 \times 10^{1}$ & $1.936 \times 10^{-2}$ &  $4.051 \times 10^{-2}$ & $1.784 \times 10^{-3}$ & $3.468 \times 10^{-3}$
\\
  \addlinespace[1mm]     \bottomrule
    \end{tabular}
  \end{center}
  \caption{\label{table:ablation_study_voxel_level_bev_view_kitti} Ablation study in the dataset of KITTI. The code length is $18$, corresponding to the compression ratio $3.19\%$.}
\end{table*}

\mypar{Normalization networks}
To handle equivalences, the voxel-level representation also needs to capture the scale and rotation variances. Here we use normalization networks to explicitly learn the scale and rotation. Before we feed the points to the voxel-level encoder, we use multilayer perceptions to lean the $3 \times 1$ scale vector and the $3 \times 3$ rotation matrix.  We keep those geometric information to the code and apply to the reconstructed points after the decoder. In this way,
we promote scale and rotation equivalences:  the 3D point cloud in each voxel is distributed in the unit space and has a similar orientation. The voxel-level encoder and decoder can thus focus on learning local shapes.

\mypar{Training implementations}
The encoder $\Phi_w \left(\cdot\right)$ and decoder $\Psi_w \left(\cdot\right)$ is implemented by using deep-neural-networks with trainable weights. To train the networks, we consider the following
optimization problem:
\begin{eqnarray*}
O(\ell) = & {\rm minimize}_{\Psi_w \left( \cdot \right), \Phi_w \left( \cdot \right)} &  \sum_{i=1}^M {\rm CD} \left(\Pj^{(i)}, \widehat{\Pj}^{(i)} \right),
\\ \nonumber
&{\rm subject~to~}&  \c  = \Psi_w \left( \Pj \right) , \widehat{\Pj}  =  \Phi_w \left( \c \right),
\\ \nonumber
&&  {\rm dim}(\c) \leq \ell,
\end{eqnarray*}
where  ${\rm CD} (\Pj, \widehat{\Pj})  = \sum_{j=1}^{N}  \min_{i \in \{1,  2, \cdots N \}} \left\|   \widehat{\p}_j - \p_i \right\|_2^2 /N  + \sum_{i=1}^{N}  \min_{j \in \{1,  2, \cdots N \}} \left\| \p_i  - \widehat{\p}_j  \right\|_2^2/N$ is the Chamfer distance~\cite{AchlioptasDMG:17}. An effective representation should be $\ell M \ll 3N$, $O(\ell) \rightarrow 0$.

\section{Experimental results}
\label{sec:experiments}
\mypar{Dataset} We validate the proposed PCT in a standard autonomous-driving dataset,  KITTI~\cite{GeigerLU:12}, which has been recorded from a moving platform while driving in and around Karlsruhe. Real-time LiDAR sweeps  are collected by a Velodyne HDL-64E rotating 3D laser scanner, with 10 Hz, 64 beams, 0.09 degree angular resolution, around 1.3 million points/second, 360 horizontal, 26.8 vertical field of view.

\mypar{Experimental Setup}  For each real-time sweep in KITTI, we partition the space into voxels with the size of $1 \times 1 \times 10$ meter$^3$. We train $400$ LiDAR sweeps and test $100$ LiDAR sweeps. We select training sweeps and testing sweeps from separate logsets to avoid data snooping. For GIN, $\mathcal{K} = \{1, 4, 8, 16\}$. To evaluate the performance, we compare the reconstruction based on the codes  to the original LiDAR sweep.

\begin{figure}
  \begin{center}
\includegraphics[width=0.7\columnwidth]{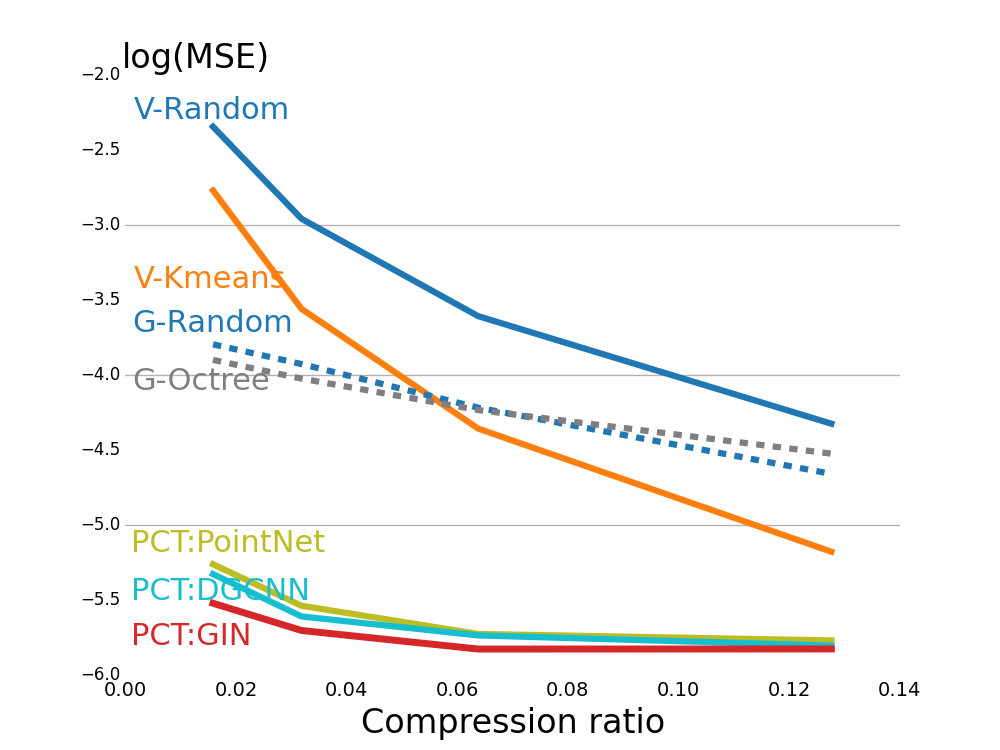}  
  \end{center}
  \caption{\label{fig:ratio_distortion}  Mean square errors as a function of the compression ratio in the dataset of KITTI.}
\end{figure}
\mypar{Results} We validate the proposed PCT from two aspects. In the ablation study, we add each component at a time to validate the effectiveness of each component; in the ratio-distortion analysis, we compare the PCT with other competitors. Due to the limited space, we only show the quantitative results for KITTI. Table~\ref{table:ablation_study_voxel_level_bev_view_kitti}
shows the ablation study in the dataset of KITTI. For the encoder, we consider three components: graph inception convolution (inception), mean-aggregation  (mean-agg), normalization (norm).  When the graph inception convolution is not checked, we consider a single $K=1$; when the mean-aggregation is not checked, we consider the maximum-aggregation; when the normalization is not checked, we do not use normalization networks. For the decoder, we consider two components: fully-connected layers and FoldingNet. We use five metrics to evaluate the performance: earth-mover distance (EMD), Chamfer distance (CD), mean square error (MSE), mean and variance~\cite{AchlioptasDMG:17}. For MSE, we consider the difference between each original 3D point and its closest correspondence in the reconstruction. For all the metrics, lower values indicate better results. We see that  each component of graph inception convolution improves the reconstruction performance; fully-connected layers consistently outperform FoldingNet.

Figure~\ref{fig:ratio_distortion} shows the mean square error as a function of compression ratios. We vary the code length in each voxel as $9, 18, 36, 72$, leading to various compression ratios. The $x$-axis is the compression ratio; the $y$-axis is the logarithm-scale mean square error.  We consider give comparison methods. Uniform resampling randomly selects a few 3D points and use this subset to represent the overall 3D point clouds. We consider resampling based on either  the entire 3D spatial space (G-Random) or the voxels (V-Random). For voxel-based resampling, we select the same number of 3D points in each voxel to promote spatial uniformity; kmeans-based representation (V-Kmeans) adaptively selects cluster centers in each voxel and we select the same number of 3D points in each voxel. Kmeans is computationally expensive and cannot afford global optimization.  These three approaches preserve information based on the spatial domain. We also consider two learning-based approaches to implement the voxel-level encoding in the framework of PCT. PointNet encodes the 3D points in each voxel into a few features by using deep neural networks~\cite{AchlioptasDMG:17}; dynamic graph convolutional neural networks (DGCNN) are the extension of PointNet by introducing a graph structure~\cite{WangSLSMS:18}. Compared to DGCNN, GIN ensures translation invariance and adopts inception structures, mean-aggregation and the normalization networks. We ensure all methods preserves the same number of data from the original sweeps. We see that PCT with GIN outperforms its competitors.  The advantage of the PCT is to transform the 3D points from the spatial domain to a feature domain, which is similar to the mechanism of the classical Fourier transform.

\section{Conclusions}
We propose the PCT to provide compact representations for large-scale 3D point clouds. The PCT includes two phases: 3D partition and voxel-level representations, which makes it acts like the blocked discrete cosine transform for 2D images. We propose GIN to improve voxel-level representations. The proposed PCT is applied to represent real-time LiDAR sweeps and significantly outperforms its competitors.

\bibliographystyle{IEEEbib}
\bibliography{bibl_jelena}

\end{document}